%% file: carmera_readay.tex
\documentclass[pmlr]{jmlr}% new name PMLR (Proceedings of Machine Learning)

\usepackage{xspace}
\usepackage{graphicx}
\usepackage{enumitem}
\usepackage{multicol}
\usepackage{multirow}
\usepackage{times}
\usepackage{latexsym}
\usepackage{kotex}
\usepackage{float}

\usepackage{longtable}% for long tables

\usepackage{booktabs}
\usepackage[load-configurations=version-1]{siunitx} % newer version

\usepackage{wrapfig,lipsum,booktabs}

\usepackage{pifont}% http://ctan.org/pkg/pifont
\usepackage{xcolor}
\usepackage{pdfrender}
\newcommand{\xmark}{\ding{55}}%
\usepackage{colortbl}
  
\newcommand*{\boldcheckmark}{%
  \textpdfrender{
    TextRenderingMode=FillStroke,
    LineWidth=.5pt, % half of the line width is outside the normal glyph
  }{\checkmark}%
}
\newcommand*{\boldxmark}{%
  \textpdfrender{
    TextRenderingMode=FillStroke,
    LineWidth=.5pt, % half of the line width is outside the normal glyph
  }{\xmark}%
}
\definecolor{forestgreen}{rgb}{0.13, 0.55, 0.13}

\newcommand{\mimicsql}{\texttt{MIMICSQL}\xspace}
\newcommand{\mimicsparql}{\texttt{MIMIC-SPARQL}\xspace}
\newcommand{\sparqlset}{\texttt{MIMIC-SPARQL*}\xspace}
\newcommand{\sqlset}{\texttt{MIMICSQL*}\xspace}
\newcommand{\accst}{$Acc_{ST}$\xspace}

\makeatletter
\def\set@curr@file#1{\def\@curr@file{#1}} %temp workaround for 2019 latex release
\makeatother

\theorembodyfont{\upshape}
\theoremheaderfont{\scshape}
\theorempostheader{:}
\theoremsep{\newline}

\jmlrvolume{149}
\jmlryear{2021}
\jmlrworkshop{Machine Learning for Healthcare}

\title[Knowledge Graph-based Question Answering with Electronic Health Records]{Knowledge Graph-based Question Answering with \\ Electronic Health Records}

\author{%
\Name{Junwoo Park} \Email{junwoo.park@kaist.ac.kr}\\
\Name{Youngwoo Cho} \Email{cyw314@kaist.ac.kr}\\
\addr KAIST / Daejeon, South Korea
\AND
\Name{Haneol Lee} \Email{studymode@yonsei.ac.kr}\\
\addr Yonsei University / Wonju, South Korea
\AND
\Name{Jaegul Choo} \Email{jchoo@kaist.ac.kr}\\
\Name{Edward Choi} \Email{edwardchoi@kaist.ac.kr}\\
\addr KAIST / Daejeon, South Korea
}

\begin{document}

\maketitle

\begin{abstract}
Question Answering (QA) is a widely-used framework for developing and evaluating an intelligent machine. In this light, QA on Electronic Health Records (EHR), namely EHR QA, can work as a crucial milestone towards developing an intelligent agent in healthcare. EHR data are typically stored in a  relational database, which can also be converted to a directed acyclic graph, allowing two approaches for EHR QA: Table-based QA and Knowledge Graph-based QA. We hypothesize that the graph-based approach is more suitable for EHR QA as graphs can represent relations between entities and values more naturally compared to tables, which essentially require JOIN operations. In this paper, we propose a graph-based EHR QA where natural language queries are converted to SPARQL instead of SQL. To validate our hypothesis, we create four EHR QA datasets (graph-based VS table-based, and simplified database schema VS original database schema), based on a table-based dataset MIMICSQL. We test both a simple Seq2Seq model and a state-of-the-art EHR QA model on all datasets where the graph-based datasets facilitated up to 34\% higher accuracy than the table-based dataset without any modification to the model architectures. Finally, all datasets are open-sourced\footnote{https://github.com/junwoopark92/mimic-sparql} to encourage further EHR QA research in both directions. 
\end{abstract}

\section{Introduction}
\begin{figure*}[t]
    \vskip -0.2in
    \begin{center}
    % \centerline{\includegraphics[height=7cm, width=17cm]{figures/table2kg_v3.pdf}}
    \centerline{\includegraphics[width=14.8cm]{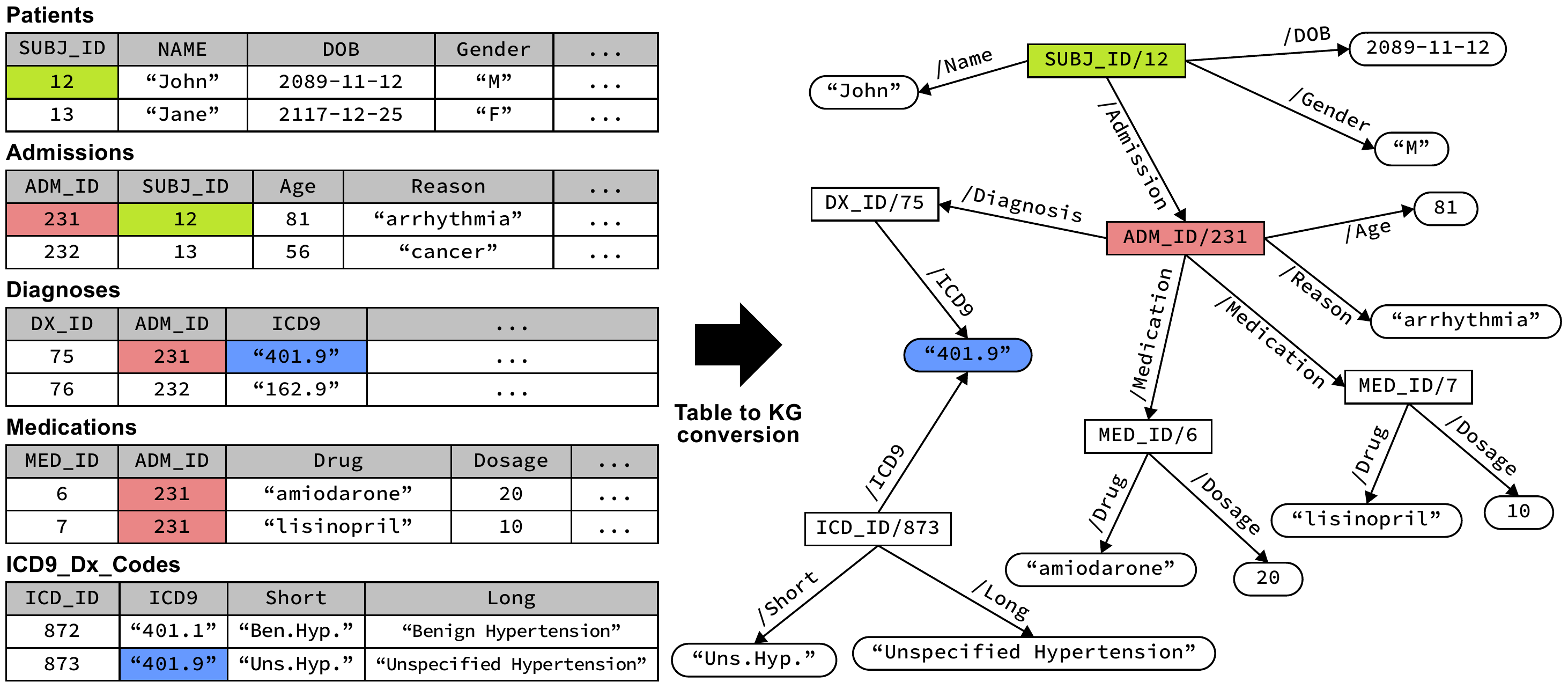}}
    \end{center}
    \vspace{-4mm}
    \caption{A subset of tables and their corresponding knowledge graph (best viewed in color). Rectangles and ellipses in the graph indicate entities and literal values, respectively. The color-coded elements are the primary key or the foreign key linking two tables.
    % Those keys linking two tables are color-coded and denoted in the graph.
    }
    \label{fig:fig1}
\end{figure*}

Electronic health records (EHR) consist of heterogeneous data (\textit{e.g.,} demographics, diagnosis, medications, and labs) stored typically in a complex relational database (RDB). An agent that can understand EHR and perform complex reasoning is not only an advancement in AI research, but also has great potential in practical applications such as clinical decision support, hospital administration, and medical chatbots.
For example, to obtain answers for questions such as \textit{``When was the last time this patient was prescribed with an antihypertensive?''} based on the information stored in the RDB, both medical experts and ordinary users typically use a graphical user interface that provides limited query assistance, or have to manually write queries such as \textit{``select count ( prescriptions.timestep ) from patients inner join admissions on patients.subject\_id = admissions.subject\_id inner join prescriptions on admissions.hadm\_id = prescriptions.hadm\_id where prescriptions.drug = antihypertensive''}.
It is laborious for both medical experts and ordinary users to write queries because in order to obtain accurate answers, they have to exactly know the hospital RDB schema and the information each table contains, and be proficient in the query language.
An intelligent EHR machine that understands the RDB information, the query language, and the natural-language questions (NLQ) can remove significant burden off both experts and ordinary users who seek information from the EHR.

Recently, the Question-Answering (QA) task has played a vital role in developing and evaluating machine intelligence \citep{bordes2014question,antol2015vqa,seo2016bidirectional,zhong2017seq2sql}, and we also aim to develop an intelligent EHR agent in a QA framework, namely EHR QA. 
As for performing QA on RDB data (\textit{i.e.,} tabular data), translating NLQ to SQL (\textit{i.e.,} NLQ2SQL) has become the main approach. This trend can also be seen in EHR QA, as \citeauthor{wang2020text} released the first NLQ2SQL dataset for EHR, namely \textit{MIMICSQL}.
% \ecedit{This trend can also be seen in EHR QA, as Wang et al (properly cite) released the first NLQ2SQL dataset for EHR, namely \textit{MIMICSQL}.}
The RDB, however, stores data typically in a fragmented manner via complex schemas, in order to maximize the efficiency in large-scale data processing and storage.
Therefore, SQL has functions such as \textit{JOIN} to aggregate information scattered across multiple tables.
This characteristic leads to decreased QA performance of the model trained on NLQ-SQL pairs when the SQL samples are generated from complex schemas.

Although EHR data are typically stored in a relational database, it can be seen as a Directed Acyclic Graph (DAG), or more specifically, a Knowledge Graph (KG) of clinical facts (see Figure~\ref{fig:fig1}), thus allowing two approaches for EHR QA: table-based QA and graph-based QA.
A knowledge graph is a multi-relational graph composed of entities as nodes and relations among them as typed edges (\textit{e.g.,} John-/medication-amiodarone, amiodarone-/dosage-30ml).
SPARQL, which is one of query languages to extract information from KG, supports reasoning over multiple edges of the KG to infer the right answer for the given question.
Unlike SQL, SPARQL does not require special functions (\textit{e.g.,} \textit{JOIN}) to collect scattered information, as KG maintains relevant information in a single graph.
We actually confirm in the experiment that SPARQL queries is closer to NLQs than SQL queries (Figure~\ref{fig:innerjoin}), and that even if the RDB schema becomes complex, the length of SPARQL queries does not increase significantly compared to that of SQL queries (Figure~\ref{fig:fig2}).
Motivated by these findings, we hypothesize that the graph-based approach could be the superior approach for EHR QA, since graphs can represent relations between entities and values more naturally compared to tables, which require \textit{JOIN} operations.
Despite such potential of NLQ2SPARQL, to our knowledge, there has not been any prior study exploring the idea of NLQ2SPARQL for EHR QA, and no NLQ-SPARQL pair datasets based on EHR. The goal and contribution of this work can be summarized as follows:

\begin{itemize}[leftmargin=*]
\item For an empirical comparison between the two, we construct both a table-based and a graph-based EHR QA datasets based on \mimicsql~\citep{wang2020text}, an existing table-based EHR QA dataset. We modify \mimicsql to follow the original schema of MIMIC-III~\citep{johnson2016mimic}, thus building \sqlset and its graph-based counterpart \sparqlset.
\item We test the state-of-the-art EHR QA model called TREQS~\citep{wang2020text}, on both \sqlset and \sparqlset where TREQS gained a 5.1\% boost in predicting the correct relations between entities, and a 3.6\% boost in predicting the correct answer.
\item We carefully analyze the reason for the graph-based method's superiority over the table-based method, presenting histograms in which the overall length of SQLs increases when the RDB schema becomes complex, and an experiment in which the number of correct SQLs decreases when the number of \textit{JOIN}s increases compared to SPARQL.
\item To the best of our knowledge, this is the first work to propose and empirically demonstrate the graph-based EHR QA.
In order to encourage further EHR QA research in both the table-based and graph-based directions, we open-source both \sqlset and \sparqlset.
\end{itemize}

\subsection*{Generalizable Insights about Machine Learning in the Context of Healthcare}
This work is closely related to healthcare from three perspectives. First, the development of an agent that can understand EHR and interact with humans in natural language can provide various conveniences to healthcare workers and consumers. Second, through the task of QA, we can quantitatively evaluate how well the machine understands medical data.
Lastly, by extending the table QA dataset to the graph QA dataset, we can explore what is more suitable data structure for performing QA on the complex EHR.

% This section is \emph{required}, must keep the above title, and should
% be the final part of your introduction.  In about one paragraph, or
% 2-4 bullet points, explain what we should \emph{learn} from reading
% this paper that might be relevant to other machine learning in health
% endeavors.

% For example, a work that simply applies a bunch of existing algorithms
% to a new domain may be useful clinically but doesn't increase our
% understanding of the machine learning and healthcare; if that study
% also investigates \emph{why} different approaches have different
% performance, that might get us excited!  A more theoretical machine
% learning work may be in how it enables a new kind of clinical study.
% \emph{Reviewers and readers will look to evaluate (a) the significance
%   of your claimed insights and (b) evidence you provide later in the
%   work of you achieving that contribution}

\section{Background}
\subsection{Electronic Health Records QA}
QA is a discipline in the fields of information retrieval and natural language processing, concerned with building systems that automatically answer questions posed by humans in natural language. In the medical domain especially, EHR QA is a research area with great potential for real-world impact, as it provides an interface where both medical experts and ordinary users can easily query clinical facts and statistics via natural language.

Recent QA research can generally be categorized into unstructured and structured QAs.
The former mainly handles free-text, both unimodal (\textit{e.g.,} machine reading comprehension (MRC) datasets \citep{rajpurkar2016squad, nguyen2016ms}) and multi-modal (\textit{e.g.,} Visual Question Answering (VQA) \citep{antol2015vqa, johnson2017clevr}).
Structured QA can largely be divided into table-based QA \citep{neelakantan2015neural, zhong2017seq2sql} and graph-based QA \citep{berant2013semantic, yih2015semantic}. EHR generally consist of multiple sources of information such as patient demographics, diagnosis history, medication records, lab results, clinical notes, radiology images, etc.
Such heterogeneous information is usually stored in a relational database, which can also be seen as a Directed Acyclic Graph (DAG) of clinical fact, or more specifically a Knowledge Graph (see, for example, Figure~\ref{fig:fig1}).
Therefore, we view EHR QA as a type of structured QA, allowing both a table-based and a graph-based approaches.

Previously, \citeauthor{pampari2018emrqa} constructed \texttt{emrQA}, an MRC dataset based on clinical notes.\texttt{emrQA}, however, focuses only on the free-text notes, thus not suitable for training machines to leverage the structural properties of EHR.
Recently, \citep{wang2020text} used MIMIC-III \citep{johnson2016mimic}, a publicly available EHR dataset, to construct \mimicsql, a table-based (\textit{i.e.,} text-to-SQL) QA dataset.
However, while constructing \mimicsql, they preprocessed the tables of MIMIC-III such that the original schema was heavily modified.
Therefore the final dataset (questions and tables) does not reflect the complex, yet interesting structure of MIMIC-III, thus losing real-world applicability and the value as a tool for fostering machine intelligence.

\subsection{Question-to-Query Generation}
To solve a structured QA, the widely used approach is a Natural Language Question-to-Query (NLQ2Query) generation, which is a sub-task of semantic parsing. The semantic parsing aims at translating a natural language text to a logical form, which is a formal semantic representation of the given natural text. For various applications, it is desirable for the logical forms to be immediately executable by the applications, such as SQL queries in the database and SPARQL queries in the knowledge graph.

Recently, NLQ2SQL generation has attracted significant attention and demonstrated successful results in various applications, including WikiSQL~\citep{zhong2017seq2sql, xu2017sqlnet} for Wikipedia, GeoQuery~\citep{catherine2018improving} for US geograpy and Spider~\citep{Tao2018spider} for domain generalization. 
In the literature of NLQ2SQL, the standard approach is the sequence-to-sequence (Seq2Seq) based method. Seq2Seq based methods \citep{sutskever2014sequence,li2016lang2log,bryan2018decathlon} have an encoder and a decoder, where both components use a sequence model such as the LSTM. More specifically, Seq2Seq based methods, which are trained with NLQ-SQL pairs, first encode input questions into vector representations and then decode the corresponding SQL queries conditioned on the encoded vectors. 

Naturally, these successes led to the work~\citep{wang2020text} in the medical field with abundant EHR data. Note that the wide use of medical jargons and abbreviations create significant challenge, making it difficult to translate natural questions to the SQL form. They cause various problems including difficult decoding and the Out-of-Vocabulary (OOV), where the model has to translate words that are unseen in training dataset. To deal with the OOV problem, a pointer generating mechanism~\citep{see2017get} is added to the Seq2Seq methods, where the decoder generates placeholder tokens and copies a word from the input question to substitute the placeholder token.
TREQS is the state-of-the-art model proposed by \citeauthor{wang2020text} for EHR QA, which uses the pointer generating mechanism and two attention mechanisms: the temporal attention on questions, and the dynamic attention on SQL, and attentive-copy techniques. This model is able to handle the unique challenges in EHR QA and is robust to the unseen questions. TREQS demonstrated its effectiveness on the EHR NLQ2SQL dataset (\mimicsql).

Compared to the table-based QA where information is stored in tables, we can think of graph-based QA where information is stored in knowledge graphs (KG). Recently, QA over KG (KGQA) is actively studied and also cast into the semantic parsing ~\citep{hegolub2016character,liang2016neural,yin2018structvae,cheng2018weakly,NIPS2018daitoact,saha2019complex}. As SQL is used for retrieving information from tables, SPARQL is a popular query language used to retrieve information in a KG. Then naturally, by modifying the NLQ2SQL framework to NLQ2SPARQL, we can tackle the QA over KG using the well-established Seq2Seq based methods. SPARQL queries, which does not involve collecting scattered information (\textit{i.e.,} \textit{JOIN} in SQL), are more concise and shorter than SQL queries for the same natural language questions. 
This is advantageous for query generation because there is less possibility of error propagation in Seq2Seq framework with autoregressive characteristics. Thus, we assume that SPARQL may be the more suitable logical form than SQL for the NLQ2Query task. To demonstrate this assumption, we construct NLQ-SPARQL pairs and compare the statics of the SPARQL to that of SQL. Actually, as shown in Figure~\ref{fig:innerjoin}, when the number of \textit{JOIN}s increases, which suggests the difficulty of the question, we see that the average length of SPARQL per \textit{JOIN} is similar to that of the natural language question compared to SQL.
Furthermore, we evaluate the state-of-the-art EHR QA model TREQS in both NLQ2SQL and NLQ2SPARQL settings where the model showed consistent performance increase in the latter.

\section{Dataset}

To the best of our knowledge, there is no existing dataset for graph-based EHR QA (\textit{i.e.,} NLQ2SPARQL).
In this section, we address the schema issue of \mimicsql and create \sqlset that preserves the true structure of MIMIC-III.
% Then we describe the how \sparqlset is constructed from \sqlset.

\input{tables/dataset_stat.tex}

\subsection{MIMICSQL to MIMICSQL*}
\label{dataset:mimicsqltostar}

\begin{figure*}[ht]
    \centering
    \includegraphics[width=14.8cm,trim={0cm 0.25cm 0cm 0cm},clip]{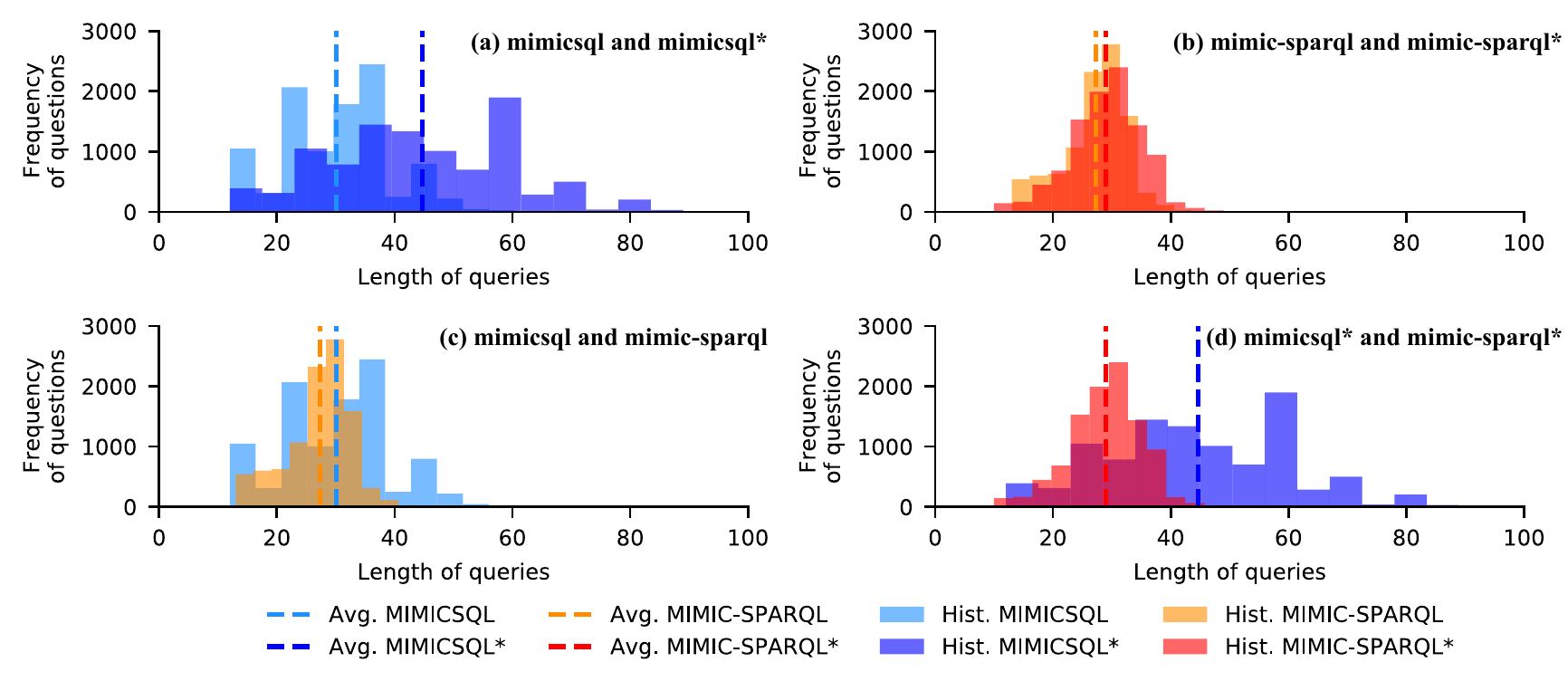}
    \caption{Histograms of query length for \mimicsql, \mimicsparql, \sqlset and, \sparqlset. We compared the histograms of each dataset for four cases. (a) and (b): When the schema becomes complex while following the original schema of MIMIC-III, the length of the query in both SQL and SPARQL increases. (c) and (d): When comparing SPARQL and SQL between datasets of the same level, the length of the SQL is increased significantly compared to that of SPARQL.}
    \label{fig:distribution}
\end{figure*}

\citeauthor{wang2020text} chose a subset of information from MIMIC-III; among the total twenty six tables, they chose nine (Patients, Admissions, Diagnoses, Prescriptions, Procedures, Labs, Diagnosis Descriptions, Procedure Descriptions, and Lab Descriptions). Then they merged those nine tables into five (Demographics, Diagnoses, Procedures, Prescriptions, and Labs) in order to simplify the information structure, which leads to simplified SQL structure.
By doing so, they compromised database normalization, such as \textit{JOIN}ing Diagnoses with Diagnosis Descriptions, thus causing duplicate column values in multiple rows.
To collect NLQ-SQL pairs, \citet{wang2020text} first define NLQ-SQL templates in the same manner with \citet{pampari2018emrqa}; they automatically generate the complete NLQ-SQL pairs (template questions), by randomly sampling the condition values from the five tables. However, these template-based NLQ-SQL pairs have some drawbacks such as not being linguistically diverse as natural questions asked by humans, and being grammatically awkward or inaccurate.
To generate more natural questions, the authors refine template questions into natural questions that a medical expert might ask by employing human annotators with clinical knowledge. Consequently, as shown in Table~\ref{tab:stats}, the natural questions has more diverse vocabulary than the template questions, while having a slightly shorter average question length. 
These statistics indicate that for the semantically same questions, the natural versions are a bit less descriptive than the template versions, while the expression of natural versions is more diverse. Both the template version and the natural version consist of $8,000$ samples for training, $1,000$ for validation, and $1,000$ for testing.

Although \mimicsql can serve as a basic clinical QA dataset, their tables are unnormalized and dissimilar from the tables used in actual hospitals.
In other words, a machine trained on \mimicsql would not generalize well to an actual hospital environment, since it was trained in an unrealistic problem space.
Therefore, to preserve the schema of MIMIC-III, we normalize \mimicsql tables back to the original nine tables, following the original schema. We also modified the SQL part of \mimicsql's NLQ-SQL pairs, such that the new tables and columns were correctly reflected.
For example, in \mimicsql, a patient's name and age are stored in the Demographic table.
However, the original schema requires a JOIN operation to find a patient's age by their names, since they are stored in two separate tables, namely Admissions and Patients respectively.
This modification increased the average number of tokens of SQL from $21.04$ to $44.68$, due to the added JOIN operations.
We denote this modified dataset as \sqlset.

\subsection{Table to Knowledge Graph}
Knowledge Graphs (KG) are a set of triples that describe the relationship between two entities or between an entity and a literal value \citep{chakraborty2019introduction}.
To extract triples from the nine tables of \sqlset, we defined a simple algorithm to map each column to either an entity, a literal, or a relation.
The primary or foreign key columns of a table are defined as entities, and non-key columns (\textit{i.e.,} property columns) are defined as literals.
For example, in the Patients table of Figure~\ref{fig:fig1}, the values of SUBJ\_ID are defined as entities that can have child nodes.
The values of Name, DOB (Date-of-Birth), and Gender are defined as literals that cannot have child nodes.
The column names of the non-key columns become relations, \textit{e.g.,} /Name links between SUBJ\_ID/12 and ``John''.
Lastly, the relation between a primary key and a foreign key is derived from the table name (\textit{e.g.,} /Admission links between SUBJ\_ID/12 and ADM\_ID/231). The final KG consists of 173,096 triples and has a max depth of five.

\begin{figure*}[ht]
    \centering
    \includegraphics[width=14.8cm]{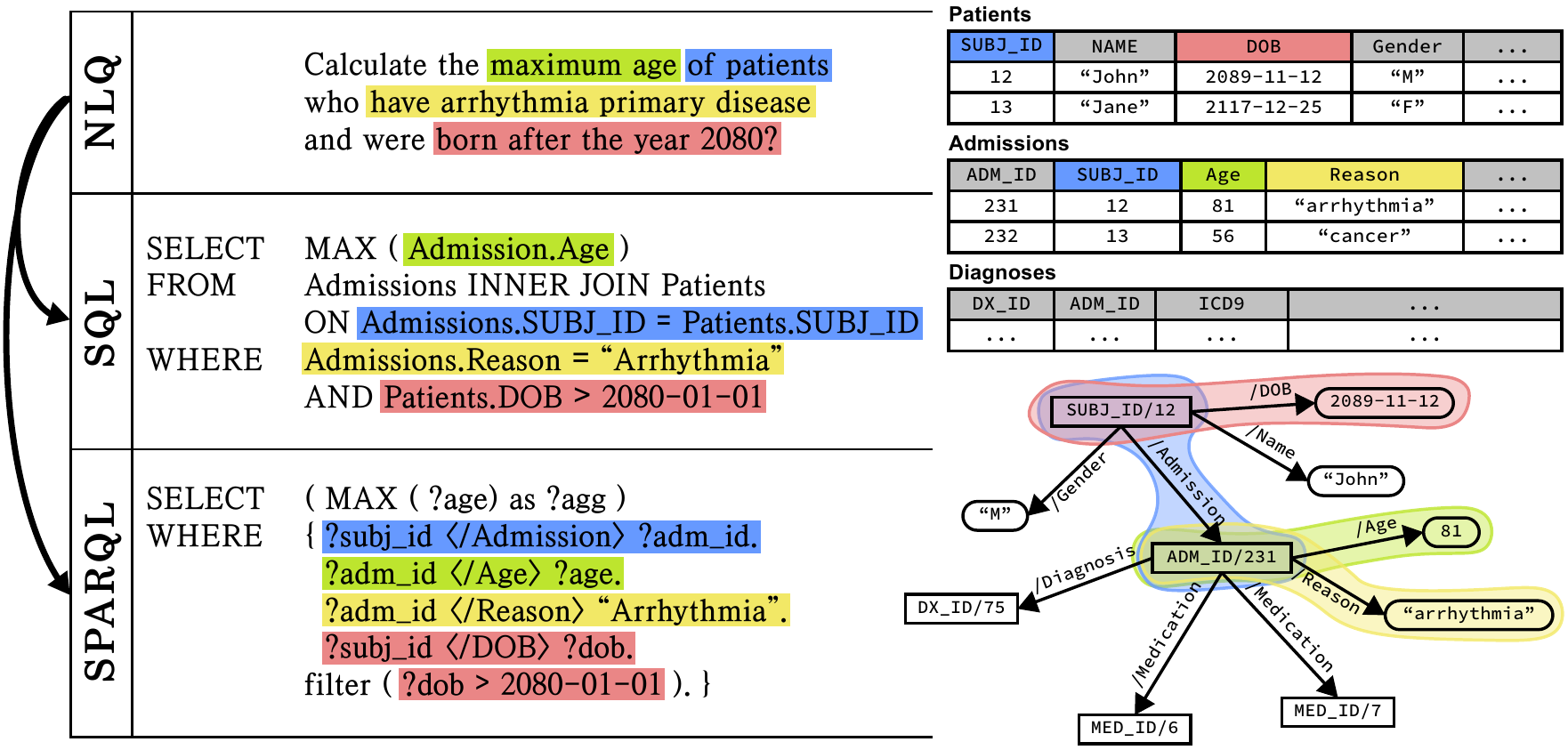}
    \caption{SQL \& SPARQL Pair: The highlighted parts in SQL and SPARQL correspond to each other. SQL requires explicit JOIN operations when linking information across tables, whereas SPARQL just hops between nodes.}
    \label{fig:fig2}
\end{figure*}

\subsection{SQL to SPARQL}
\label{sec:sqltosparl}
In order to create NLQ-SPARQL pairs from NLQ-SQL pairs, we treat the \textit{SELECT} clause and the \textit{WHERE} clause of SPARQL separately.
First, we copy the aggregation function (\textit{e.g.,} MAX, AVERAGE) and the columns from the \textit{SELECT} part of SQL, and fill the SPARQL template.
Then we extract the condition columns, comparison operator, and condition values from the \textit{WHERE} part of SQL.
In order to fill the \textit{WHERE} clause of SPARQL, which consists of triples between variables, 
we find the shortest distance between the variable in the KG\footnote{Since we know the schema of MIMIC-III, we can create a relation graph and apply Dijkstra's algorithm.}.
As for combining comparison operators (\textit{e.g.,} $>$, $<$, $=$) with triples, equal operators are handled by replacing the variable with values, whereas all other operators are handled by adding filter statements.
An example of SQL-to-SPARQL conversion is depicted by Figure~\ref{fig:fig2}.
After the conversion, we checked the semantic equivalence of SQLs and SPARQLs by querying them to the tables and the KG, where we confirmed a 100\% match.
The KG and SPARQL queries comprise the graph-based EHR-QA dataset \sparqlset.
Note that we also constructed \mimicsparql based on the five tables of \mimicsql for further empirical analysis.
Table~\ref{tab:stats} describes the basic statistics of \sqlset, \sparqlset, \mimicsql, and \mimicsparql.

\section{Experiments}

This section describes the setting and results of experiments to demonstrate the superior EHR QA performance of the graph-based approach compared to the table-based approach. 
All datasets and codes will be open-sourced after the review period.

\subsection{Experiments Setting}

\vspace{1mm}
\noindent \textbf{4.1.1 Comparison Methods. \enskip} We employed two encoder-decoder based approaches: Seq2Seq model \citep{luong2015effective} and TREQS \citep{wang2020text}.
Seq2Seq uses a bidirectional LSTM encoder and an LSTM decoder with an attention mechanism, but it cannot handle out-of-vocabulary (OOV) tokens that frequently occur in medical data.
Proposed with \texttt{MIMICSQL}, TREQS is a state-of-the-art NLQ2SQL translation model for clinical questions, equipped with the copying mechanism \citep{see2017get} to address the OOV issue, and the attention mechanism \citep{bahdanau2014neural} to help SQL token generation.

 The LSTM-based Seq2Seq model was implemented with PyTorch.
 As for TREQS, we used the official code released by the original authors, rather than implementing from scratch.
 In order to carefully compare the grammatical difference between two query languages without modifying the model architecture, we utilized the official code\footnote{https://github.com/wangpinggl/TREQS} of TREQS and hyperparameters described in \citep{wang2020text}.
 We only tuned hyperparameters that are irrelevant to the model size. 
%  with the same methods.
%  \ecedit{What does "with the same methods" mean?}
 The number of trainable parameters of TREQS is $2,795,907$ and Seq2Seq is $3,844,096$ throughout all experiments.
As shown in Table~\ref{table:hyperparam}, the hyperparameters are chosen to optimize the model for the dataset without increasing the model size and sampled from uniform distribution.
We train all models from scratch to 20 epochs. The development set is utilized to select the best configuration. In the test time, we employ a beam search algorithm to generate the output query sentences where the beam size is set to 5.
Training both models required an average of one hour on a TITAN-Xp GPU.

\vspace{1mm}
\noindent \textbf{4.1.2 Dataset Descriptions. \enskip}  
The datasets used in the experiment are grouped according to three criteria.
First, the QA datasets are categorized into NLQ-SQL pairs (\mimicsql and \sqlset) and NLQ-SPARQL pairs (\mimicsparql and \sparqlset) depending on the type of a logical form.
Note that NLQs are the same for all datasets, and therefore any performance difference is solely caused by whether we use the graph-based or table-based approach.
The second criterion is whether to follow the original schema of MIMIC-III. \mimicsql is constructed with a simplified MIMIC-III schema, whereas \sqlset maintains the original schema of MIMIC-III. 
Since the knowledge graph is created by converting the MIMIC-III tables, \mimicsparql and \sparqlset were converted from their respective counterparts, \mimicsql and \sqlset. Lastly, there are two versions of NLQ, namely template questions and natural language questions. As described in Section~\ref{dataset:mimicsqltostar}, the NLQ-SQL pairs in MIMICSQL were constructed based on human-defined SQL templates. As seen in Table~\ref{tab:stats}, the template questions are less diverse compared to the natural questions in terms of linguistic expression, making it easier for the model to understand the intent of the question. While natural questions are more similar to human language than template questions are, it is generally more difficult for the model to understand the intent.

To demonstrate the performance between SPARQL and SQL, we compare the performance of models trained with \mimicsql against \mimicsparql, and \sqlset against \texttt{MIMIC-} \texttt{SPARQL*}. Additionally, we found an improved method to tokenize SQLs of \mimicsql. In the SQL queries provided in the original \mimicsql, the table name and the column name are not separated and are considered as a single token. This makes it difficult for the model to learn the relationship between the \textit{SELECT} clause and \textit{FROM} clause. Thus, we separate the table name from the column name. The effectiveness of this tokenization is explained in Section~\ref{exp:gapwithpaper}. 

\vspace{1mm}
\noindent \textbf{4.1.3 Evaluation Metrics. \enskip} 
We use the following three metrics to measure the question answering performance of the models. Table~\ref{tab:metric} shows the ground truth SQL and four SQL examples to elaborate how each metric is measured.
\begin{itemize}[leftmargin=*]
\item \textbf{Logic Form Accuracy \citep{zhong2017seq2sql} ($Acc_{LF}$)} is calculated by comparing the generated SQL/SPARQL queries with the true SQL/SPARQL queries token-by-token. This is the most strict way of measuring the model performance.
\item  \textbf{Execution Accuracy ($Acc_{EX}$)} is calculated based on the correctness of the answer retrieved by querying the tables/KG with the generated SQL/SPARQL. This is a more lenient way of measuring the model performance, since the model can generate incorrect SQL/SPARQL queries but still get the correct answer by luck.
\item \textbf{Structural Accuracy ($Acc_{ST}$)} is similar to $Acc_{LF}$, but disregards the condition value tokens (\textit{e.g.,} numeric values or string values).
%\ecedit{Furthermore, \accst allows the permutation of the AND/OR conditions as opposed to $Acc_{LF}$, where the exact order must be met.}
We use \accst to evaluate how well the model understands the relations between columns/entities, and successfully converts NLQ to SQL/SPARQL structure, with less emphasis on the exact condition values (similar to how models are evaluated for the Spider dataset~\citep{Tao2018spider}).
\end{itemize}

\input{tables/metric_example}

\subsection{Experiments Results}
\label{exp:results}

\input{tables/nl_results_table.tex}

\vspace{1mm}
\noindent \textbf{4.2.1 SQL vs SPARQL. \enskip}
As shown in Table~\ref{main_nl_exp}, regardless of the type of NLQ (template or natural), the SQL dataset (\mimicsql) shows competitive performance to its SPARQL counterpart (\texttt{MIMIC-}\texttt{SPARQL}) when using the simplified schema (\textit{i.e.,} five tables).  
When using the original MIMIC-III schema, however, the graph-based approach (\sparqlset) consistently outperforms the table-based approach (\sqlset).
It is noteworthy that the \accst gap between SQL and SPARQL dramatically increases when moving from the modified schema to the original, indicating that the models better understand the complex relations between pieces of information when trained on the SPARQL dataset.
Furthermore, while there was a significant performance drop when moving from \mimicsql to \sqlset, especially when using a simpler model (\textit{i.e.,} Seq2Seq), there was either marginal decrease or even slight increase in performance when moving from \mimicsparql to \sparqlset, indicating the effectiveness of the graph-based approach as the knowledge structure becomes more complex.

These results support our hypothesis that the graph-based approach could be more suitable for EHR QA as graphs can represent relations between entities and values more concisely compared to tables, which require \textit{JOIN} operations.
Specifically, the statistics of the dataset in Table~\ref{tab:stats} show that the query length of \sqlset is nearly twice as long as that of \sparqlset. This is because a single \textit{JOIN} in SQL requires 11 tokens (including ‘=’ and ‘.’), while a single hop in SPARQL requires only 3 (\textit{i.e.,} subject, predicate, and object).
This characteristic between SQL and SPARQL is also explicitly visualized in Figure~\ref{fig:distribution}, where we can readily observe the drastic change of SQL query length when moving from simple to complex schema, whereas SPARQL query length stays relatively constant.
Additionally, unlike SPARQL, SQL requires the model to learn the hierarchy between a table and its columns, in addition to the relations between tables. 
Such syntactic difference between SQL and SPARQL originates from the intrinsic difference between the relational tables and a knowledge graph in terms of how to connect information (joining multiple tables versus hopping across triples).
This leads to the superior performance of the graph-based approach over the table-based one especially in \accst, which only evaluates the structural accuracy (\textit{i.e.,} disregarding the correctness of the condition values).

\begin{figure}[ht]
    \centering
    \includegraphics[height=6cm, width=14.8cm]{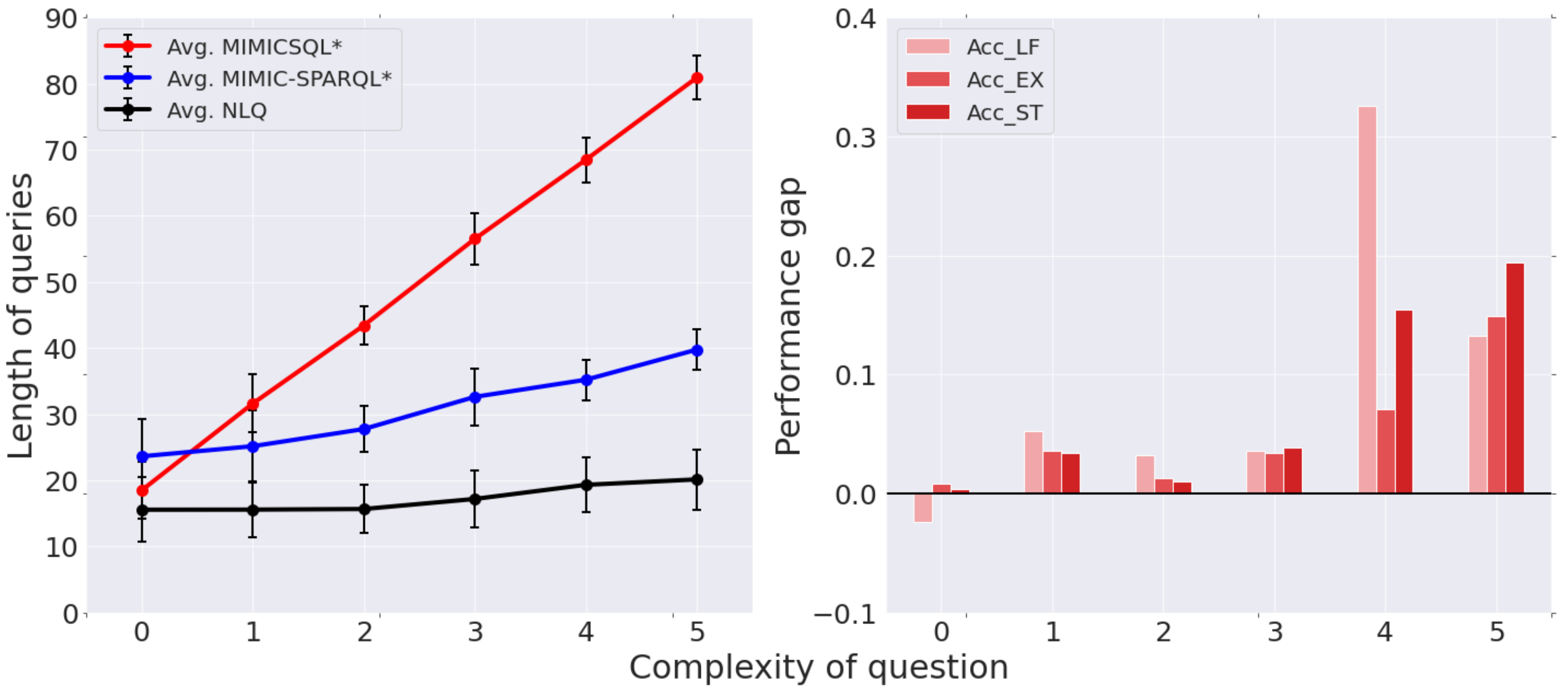}
    \caption{\textit{Left:} When the complexity of the queries increases, the length of all queries increases. The length of SPARQLs increases with a slope more similar to that of NLQs than that of SQL queries. \textit{Right:} This leads to the increased performance gap between graph-based and table-based models for complex questions.}
    
    % \caption{Query length per JOINs: When the number of JOINs included in the SQL queries increases, the length of all queries increases. As shown here, the length of SPARQL queries increases with a slope more similar to that of NLQs than that of SQL queries.}
    \label{fig:innerjoin}
\end{figure}

\vspace{1mm}
\noindent \textbf{4.2.2 Analysis of NLQ type: Template and Natural. \enskip}
As mentioned in Section~\ref{dataset:mimicsqltostar}, MIMICSQL provides both template questions and natural questions, thus giving us the option to either convert template questions to SQL/SPARQL, or convert natural questions to SQL/SPARQL.
As the vocabulary sizes in Table~\ref{tab:stats} indicates, the linguistic patterns of template questions are less diverse than the patterns of natural questions because template questions are generated by sampling only the condition values based on a predefined template to create the NLQ2Query pairs.
On the other hand, the natural questions contain various phrases thanks to the refinement process by human annotators.
Naturally, the more diverse the question patterns are, the harder the task is for the model to correctly encode the given questions to a latent space.
As shown in Table~\ref{main_nl_exp}, training the model on template questions demonstrates superior performance compared to training the model on natural questions, regardless of whether we use the table-based approach or the graph-based approach.
It is noteworthy that the graph-based approach demonstrated approximately 30\% higher \accst than the table-based approach. 
Moreover, in the natural questions, regardless of the logical form and model, all models lose performance when the schema becomes complex from five tables to nine tables. However, in the template questions, TREQS trained with SPARQL maintains the performance even if the schema becomes complex.
This indicates that the graph-based approach is robust to the complexity of the schema when the questions are straightforward enough to encode (\textit{i.e.,} less diverse vocabulary) as explained in Section~\ref{dataset:mimicsqltostar}.

 \begin{figure*}[htbp!] %[t]
    \centering
    \includegraphics[width=14.8cm]{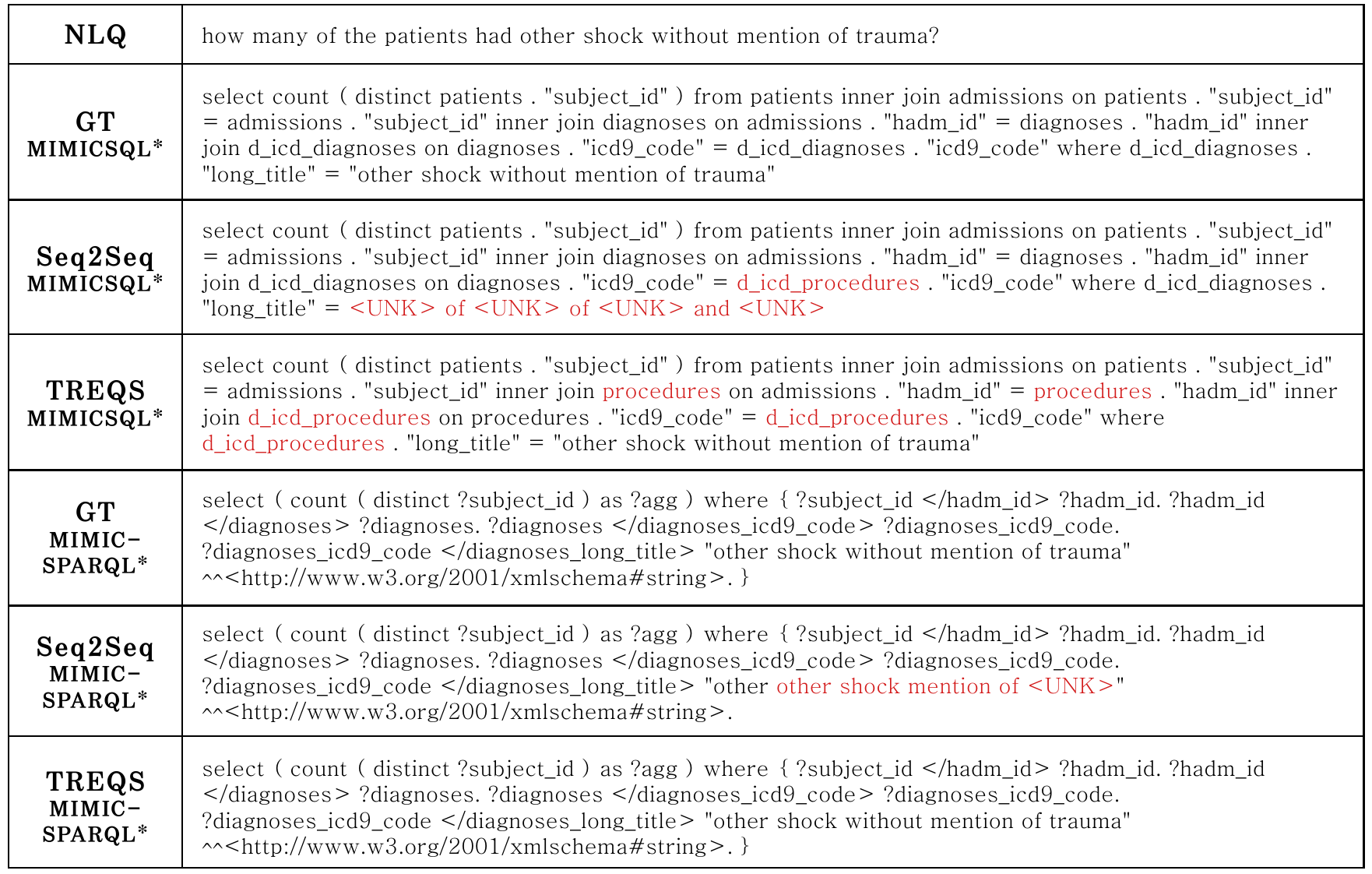}
    \caption{TREQS learned with \sparqlset correctly generated query structure and condition values. However, even though it is the same architecture, when learned with \sqlset, the model incorrectly generated the structure of SQL (highlighted in red). }
    \label{fig:fig9999}
    % Bottom: Both SQL and SPARQL were generated incorrectly, but in the case of SPARQL, the query structure was still correctly generated.
\end{figure*}

\vspace{1mm}
\noindent \textbf{4.2.3 Performance Gap by Question Complexity. \enskip}
We can reliably say that if an SQL query contains many \textit{JOIN} operations, then the corresponding NLQ involves multiple logical connections (\textit{i.e.,} reasoning) to find the answer.
Thus, we can safely assume that the complexity of the NLQ is proportional to the number of \textit{JOIN}s. As explained in Section~\ref{sec:sqltosparl}, the SPARQL datatsets share the same NLQ as the SQL datasets, and we can classify the SPARQL queries based on the number of \textit{JOIN}s in the corresponding SQL queries. In addition, from the SPARQL perspective, the number of hops between triples increases when the number of \textit{JOIN}s increases. 

We confirmed in Table~\ref{main_nl_exp} that the performance drop of table-based approaches was greater than that of graph-based when the schema became complex from the five tables to nine tables. Figure~\ref{fig:innerjoin} presents possible explanations for this phenomenon. We classified the questions according to complexity, and plotted the performance gap of the two models (\textit{i.e.} TREQS models trained with \sqlset and \sparqlset respectively), for each question group as a bar graph for the three metrics (see \textit{Right}). The performance gap between two models increases for complex questions that require more than three \textit{JOIN}s.
%We focused on changing the length of the SQL/SPARQL depending on the question complexity in order to find the reason for the discrepancy.
We assumed the different natures of SQL and SPARQL led to this performance gap, and analyzed the sequence length of SQL and SPARQL queries per varying question complexity.
% We present in Figure~\ref{fig:innerjoin} the change in the length of the query as we increase the complexity of NLQs.
As the complexity of question increases, the length of all types of queries tends to increase (see \textit{Left}).
Note that the increase in the length of NLQs is not larger than that of SQL queries.
This is because of the syntactic nature of SQL, where the multiple keywords for performing \textit{JOIN}s between tables are added. On the other hand, the SPARQL queries do not increase in length significantly compared to the SQL queries.
Through this trend, we judge that the query structure of SPARQL is closer to NLQ than SQL. 
% As a result, the models trained with the SPARQL datasets correctly generate the queries compared to the models trained with the SQL datasets for the questions that require one or more \textit{JOIN}s.

\vspace{1mm}
\noindent \textbf{4.2.4 Qualitative Comparison between Generated Queries. \enskip}
\label{exp:gapwithpaper}
We demonstrate the qualitative results to study the differences between the model, and the differences between the table-based and the graph-based datasets.
For the same NLQ, we present queries for six cases as shown in Figure~\ref{fig:fig9999}.
In terms of the model, TREQS handles OOV correctly by copying the condition value in the NLQ. However, Seq2Seq generates $<$UNK$>$ tokens for the condition values in both \sqlset and \sparqlset, since there exist several abbreviations and rarely occurring words in the EHR.
In terms of a logical form, the models that learned SQL incorrectly create ``\textit{d\_icd\_diagnoses}'', which correspond to the query structure, as ``\textit{d\_icd\_procedures}''. On the other hand, although the Seq2Seq model generated condition values incorrectly due to the OOV problem, the models that learned SPARQL generate the query structure perfectly.

\vspace{1mm}

\section{Discussion \& Limitation} 
From the vast amount of EHR data, retrieving desired information is laborious for both medical experts and ordinary users.
EHR QA is a meaningful task in that it provides a natural language interface that allows users to readily extract information. In addition, it is a promising research field that can serve telemedicine to patients in the era where the contact-free treatment becomes important. In this work, we focused on the differences in the data structure of the relational database and the knowledge graph, and analyzed the variation of logical forms in complex schema.
Motivated by these findings, we hypothesized that a graph-based approach is a more suitable choice for conducting complex EHR QA than a table-based approach.
With extensive experiments, we empirically demonstrated the superior performance of the graph-based approach.
To encourage EHR QA research in both directions, we constructed a pair of EHR QA datasets, one table-based (\sqlset) and another graph-based (\sparqlset).
To the best of our knowledge, this is the first work to propose and successfully demonstrate the graph-based EHR QA.

\paragraph{Limitations}
The graph-based approach, however, has its own limitation in terms of inference time compared to the table-based approach.
%When the size of the graph grows (\textit{e.g.} converting massive EHR data to graphs), it requires considerable time to process SPARQL queries, hence the slow response.
Note that the SQL processing time scales favorably as the table size grows, compared to the SPARQL processing time as the graph size grows (especially in terms of the number of edges).
In this work, the average response time for both the table-based (SQL) and the graph-based (SPARQL) approaches were less than 1 second, since both datasets were constructed with a subset of MIMIC-III.
But if we were to convert the entire MIMIC-III dataset to a knowledge graph, then scalability might become a bottleneck when deploying the model in practice (\textit{e.g.} SPARQL taking several seconds to retrieve an answer).
Note that, however, this is strictly related to the inference phase, and not the training phase. The model training is actually faster with SPARQL, since SPARQL queries are shorter than SQL queries on average.
As future work, we plan to extend both our datasets (\sqlset and \sparqlset) to cover a larger subset of MIMIC-III and to address the scalability of the graph-based approach.

\section*{Acknowledgements}
This work was supported by Institute of Information \& communications Technology Planning \& Evaluation (IITP) grant (No. 2019-0-00075, Artificial Intelligence Graduate School Program (KAIST) and No. 2020-0-00368, A Neural-Symbolic Model for Knowledge Acquisition and Inference Techniques) and National Research Foundation of Korea (NRF) grant (NRF-\\2020H1D3A2A03100945), funded by the Korea government (MSIT).

% ACKNOWLEDGEMENTS ONLY GO IN THE CAMERA-READY, NOT THE SUBMISSION
% \acks{Many thanks to all collaborators and funders!}

\bibliography{reference.bib}
\clearpage

\appendix
\section*{Appendix A.}
\input{tables/hyperparameters}
\subsection{Performance Gap between \mimicsql and \mimicsql (paper)}
\label{apen:gap}
As shown in bottom Table~\ref{main_nl_exp}, there is performance gap between the reproduced results and the reported results in original paper~\citep{wang2020text}. In the queries provided by the original \mimicsql, the table name and the column name are treated as a single token when being fed to the model.
This makes it difficult for the model to learn the relationship between the \textit{SELECT} clause and \textit{FROM} clause.
Thus, we separate the table name from the column name and report the effectiveness of this tokenization. Table~\ref{main_nl_exp} shows that training models on \mimicsql increases performance for all metrics compared to training models on \mimicsql (paper).
By simply changing the tokenization method, the execution accuracy increased by 22\% and 19\% in the Seq2Seq and TREQS that are trained with the natural questions, respectively.
\end{document}

%% file: tables/dataset_stat.tex
\begin{table}
\caption{\small Basic statistics of \sqlset and \sparqlset. The values in the parentheses are from the original \mimicsql and its counterpart \mimicsparql.}
\label{tab:stats}
\centering
\small
\begin{tabular}{lccc}
\toprule
 &  \sqlset & \sparqlset \\
%  & \begin{tabular}{@{}c@{}}\texttt{MIMIC} \\ \texttt{SQL*}\end{tabular} & 
% \begin{tabular}{@{}c@{}}\texttt{MIMIC-} \\ \texttt{SPARQL*}\end{tabular}\\
\midrule
\# of patients & \multicolumn{2}{c}{100}\\
\# of Question-Query pairs & \multicolumn{2}{c}{10,000} \\
Avg. TP question length & \multicolumn{2}{c}{18.40} \\
Avg. NL question length & \multicolumn{2}{c}{16.45} \\
\# of unique words in TP questions & 6,696 (6,693) & 6,525 (6,525)\\
\# of unique words in NL questions & 7,331 (7,329) & 7,160 (7,160) \\
%\hline
Avg. SQL/SPARQL length & 44,68 (21.04) & 28.98 (27.28) \\
Max \# of \textit{INNER JOIN}s & 5 (2) & - \\
Max depth of graph  & - & 5 (3) \\
\# of triples & -  & 173,096 (257,187) \\
\bottomrule
\end{tabular}
\end{table}

%% file: tables/metric_example.tex
\begin{table}
\caption{Four pseudo-SQL examples of measuring $Acc_{LF}$, $Acc_{EX}$, and $Acc_{ST}$}
\label{tab:metric}
\vspace{3mm}
\centering
\small
\scalebox{0.95}{
\begin{tabular}{c|l|ccc}
\toprule
%  &  SQL & LF & EX & ST  \\
% \midrule
\textbf{Ground Truth} & \textbf{\textit{select max(age) from patients where Gender = ``F'' and DoB $>$ 2020}} & \textbf{\textit{LF}} & \textbf{\textit{EX}} & \textbf{\textit{ST}} \\
\midrule
Predicted SQL1 & \textit{select max(age) from patients where Gender =``F'' and DoB $>$ 2020} & \textcolor{forestgreen}{\boldcheckmark} & \textcolor{forestgreen}{\boldcheckmark} & \textcolor{forestgreen}{\boldcheckmark} \\
Predicted SQL2 & \textit{select max(age) from patients where DoB $>$ 2020 and Gender = ``F''} & \textcolor{red}{\boldxmark} & \textcolor{forestgreen}{\boldcheckmark} & \textcolor{forestgreen}{\boldcheckmark} \\
Predicted SQL3 & \textit{select max(age) from patients where DoB $>$ 2021 and Gender = ``M''} & \textcolor{red}{\boldxmark} & \textcolor{red}{\boldxmark} & \textcolor{forestgreen}{\boldcheckmark} \\
Predicted SQL4 & \textit{select max(age) from patients where DoB $>$ 2021 and Diagnosis = ``F''} & \textcolor{red}{\boldxmark} & \textcolor{red}{\boldxmark} & \textcolor{red}{\boldxmark} \\
\bottomrule
\end{tabular}
}
\end{table}

%% file: tables/nl_results_table.tex
\begin{table*}[t]

\caption{\label{main_nl_exp} Medical QA performance on template (Top) and natural (Bottom) questions evaluated with logic form accuracy ($Acc_{LF}$), execution accuracy ($Acc_{EX}$) and the structural accuracy (\accst). In addition to \mimicsql, we include \mimicsql (Paper), the performance reported in \cite{wang2020text} and explained in Appendix~\ref{apen:gap}}

\centering

\scalebox{0.61}{
\begin{tabular}{llccccccc}
\toprule
\multicolumn{1}{c}{\multirow{2}{*}{\textbf{Method}}} &
% \multicolumn{1}{c}{\multirow{2}{*}{\textbf{Dataset}}} &
\multicolumn{1}{c}{\textbf{Dataset}} &
\multicolumn{1}{c}{\multirow{2}{*}{\textbf{\# of Tables}}} &
\multicolumn{3}{c}{\textbf{Development}} & \multicolumn{3}{c}{\textbf{Testing}} \\
 & \multicolumn{1}{c}{Template Question} & & $Acc_{LF}$ & $Acc_{EX}$ & \accst & $Acc_{LF}$ &  $Acc_{EX}$ & \accst\\
\midrule
\multirow{5}{*}{Seq2Seq}
& \mimicsql (Paper)& \multirow{4}{*}{5} & 0.098 & 0.372 & - & 0.160 & 0.323 & - \\
& \mimicsql &  & 0.327 $(\pm 0.118)$ & \textbf{0.429} $(\pm 0.075)$ & 0.467 $(\pm 0.028)$ & 0.410 $(\pm 0.120)$ & 0.455 $(\pm 0.083)$ & 0.546 $(\pm 0.022)$ \\
& \mimicsparql &  & \textbf{0.337} $(\pm 0.117)$ & 0.383 $(\pm 0.073)$ & \textbf{0.877} $(\pm 0.029)$ & \textbf{0.426} $(\pm 0.120)$ &\textbf{0.463} $(\pm 0.089)$ & \textbf{0.892} $(\pm 0.016)$ \\
\cmidrule(lr){2-9}

& \sqlset & \multirow{2}{*}{9} & 0.298 $(\pm 0.156)$ & \textbf{0.407} $(\pm 0.099)$ & 0.455 $(\pm 0.017)$ & 0.348 $(\pm 0.158)$ & 0.418 $(\pm 0.104)$ & 0.512 $(\pm 0.020)$ \\
& \sparqlset &  & \textbf{0.333} $(\pm 0.134)$ & 0.380 $(\pm 0.088)$ & \textbf{0.874} $(\pm 0.022)$ & \textbf{0.415} $(\pm 0.127)$ & \textbf{0.461} $(\pm 0.086)$ & \textbf{0.858} $(\pm 0.015)$ \\

 \midrule
\multirow{5}{*}{TREQS}

& \mimicsql (Paper)& \multirow{4}{*}{5} & 0.712 & 0.803 & -  & 0.802 & 0.825 & -  \\
& \mimicsql & & \textbf{0.752} $(\pm 0.025)$ & 0.810 $(\pm 0.012)$ & 0.982 $(\pm 0.057)$ & \textbf{0.823} $(\pm 0.038)$ & \textbf{0.843} $(\pm 0.038)$ & 0.986 $(\pm 0.082)$ \\
& \mimicsparql & & 0.735 $(\pm 0.017)$ & \textbf{0.816} $(\pm 0.014)$ & \textbf{0.989} $(\pm 0.008)$ & 0.805 $(\pm 0.010)$ & 0.841 $(\pm 0.012)$ & \textbf{0.990} $(\pm 0.007)$ \\

\cmidrule(lr){2-9}
& \sqlset& \multirow{2}{*}{9}  & 0.672 $(\pm 0.053)$ & 0.789 $(\pm 0.027)$ & 0.921 $(\pm 0.011)$ & 0.749 $(\pm 0.047)$ & 0.816 $(\pm 0.032)$ & 0.952 $(\pm 0.015)$ \\
& \sparqlset& & \textbf{0.745} $(\pm 0.007)$ & \textbf{0.824} $(\pm 0.007)$ & \textbf{0.994} $(\pm 0.001)$ & \textbf{0.814} $(\pm 0.010)$ & \textbf{0.848} $(\pm 0.007)$ & \textbf{0.991} $(\pm 0.006)$ \\

\bottomrule
\end{tabular}
}

\vspace{5mm}

\scalebox{0.61}{
\begin{tabular}{llccccccc}
\toprule
\multicolumn{1}{c}{\multirow{2}{*}{\textbf{Method}}} &
% \multicolumn{1}{c}{\multirow{2}{*}{\textbf{Dataset}}} &
\multicolumn{1}{c}{\textbf{Dataset}} &
\multicolumn{1}{c}{\multirow{2}{*}{\textbf{\# of Tables}}} &
\multicolumn{3}{c}{\textbf{Development}} & \multicolumn{3}{c}{\textbf{Testing}} \\
 & \multicolumn{1}{c}{Natural Question} & & $Acc_{LF}$ & $Acc_{EX}$ & \accst & $Acc_{LF}$ &  $Acc_{EX}$ & \accst\\
\midrule
\multirow{5}{*}{Seq2Seq}
& \mimicsql (Paper)& \multirow{4}{*}{5} & 0.076 & 0.112 & - & 0.091 & 0.131 & - \\

& \mimicsql &  & \textbf{0.197} $(\pm 0.031)$ & \textbf{0.335} $(\pm 0.024)$ & 0.360 $(\pm 0.026)$ & \textbf{0.261} $(\pm 0.031)$ & \textbf{0.358} $(\pm 0.030)$ & 0.404 $(\pm 0.025)$ \\

& \mimicsparql &  & 0.179 $(\pm 0.093)$ & 0.272 $(\pm 0.056)$ & \textbf{0.637} $(\pm 0.019)$ & 0.252 $(\pm 0.099)$ & 0.338 $(\pm 0.065)$ & \textbf{ 0.634} $(\pm 0.016)$ \\

\cmidrule(lr){2-9}
& \sqlset & \multirow{2}{*}{9} & 0.148 $(\pm 0.040)$ & \textbf{0.301} $(\pm 0.014)$ & 0.327 $(\pm 0.012)$ & 0.181 $(\pm 0.040)$ & 0.293 $(\pm 0.030)$ & 0.347 $(\pm 0.018)$ \\
& \sparqlset &  & \textbf{0.185} $(\pm 0.132)$ & 0.285 $(\pm 0.092)$ & \textbf{0.623} $(\pm 0.039)$ & \textbf{0.225} $(\pm 0.133)$ & \textbf{0.321} $(\pm 0.094)$ & \textbf{0.590} $(\pm 0.038)$ \\
 \midrule
\multirow{5}{*}{TREQS}
& \mimicsql (Paper)& \multirow{4}{*}{5} & 0.451 & 0.511 & -  & 0.486 & 0.556 & -  \\
& \mimicsql & & \textbf{0.588} $(\pm 0.005)$ & 0.707 $(\pm 0.009)$ & 0.825 $(\pm 0.005)$ & \textbf{0.661} $(\pm 0.030)$ &\textbf{0.732} $(\pm 0.024)$ & \textbf{0.821} $(\pm 0.023)$ \\

& \mimicsparql & & 0.585 $(\pm 0.006)$ & \textbf{0.721} $(\pm 0.006)$ & \textbf{0.828} $(\pm 0.008)$ & 0.649 $(\pm 0.016)$ & 0.729 $(\pm 0.012)$ & 0.818 $(\pm 0.016)$ \\
\cmidrule(lr){2-9}
& \sqlset& \multirow{2}{*}{9}  & 0.530 $(\pm 0.031)$  & 0.689 $(\pm 0.019)$  & 0.736 $(\pm 0.030)$ & 0.601 $(\pm 0.031)$ & 0.694 $(\pm 0.020)$ & 0.743 $(\pm 0.036)$ \\
& \sparqlset & & \textbf{0.577} $(\pm 0.017)$ & \textbf{0.715} $(\pm 0.013)$ & \textbf{0.818} $(\pm 0.011)$ & \textbf{0.633} $(\pm 0.010)$ & \textbf{0.722} $(\pm 0.006)$ & \textbf{0.788} $(\pm 0.010)$ \\
\bottomrule
\end{tabular}
}

\normalsize

\end{table*}

%% file: tables/hyperparameters.tex
\begin{table*}[t]
\caption{Description of hyperparameters}
% \resizebox{\columnwidth}{!}{%
% \scalebox{1.2}{
% \centering
% \small
\begin{tabular}{lllcc}

\toprule
\multicolumn{1}{c}{\multirow{2}{*}{\textbf{Method}}} &
\multicolumn{1}{c}{\multirow{2}{*}{\textbf{Dataset}}} &
\multicolumn{3}{c}{\textbf{Hyperparameters}} \\
& & Parameter & Search space &  Selected value \\
\midrule

\multirow{8}{*}{\textbf{Seq2Seq}} & \multirow{4}{*}{\sqlset} &  Batch Size & $16, 32, 48$ & $32$ \\ 
 &  &  Step decay & $[0.01, 0.8]$ & $0.1$ \\ 
 &  & Step size & $1,2,5,10$ & $2$ \\  
 &  & Learning rate & $[1\times10^{-5}, 1\times10^{-2}]$ & $0.001$  \\ 
 & \multirow{4}{*}{\sparqlset} & Batch Size & $16, 32, 48$ & $16$ \\ 
 &  &  Step decay & $[0.01, 0.8]$ & $0.8$ \\ 
 &  & Step size & $1,2,5,10$ & $2$ \\  
 &  & Learning rate & $[1\times10^{-5}, 1\times10^{-2}]$ & $0.001$  \\ 
\midrule

\multirow{8}{*}{\textbf{TREQS}} & \multirow{4}{*}{\sqlset} &  Batch Size & $16, 32, 48$ & $16$ \\ 
 &  &  Step decay & $[0.01, 0.8]$ & $0.8$ \\ 
 &  & Step size & $1,2,5,10$ & $2$ \\  
 &  & Learning rate & $[1\times10^{-5}, 1\times10^{-2}]$ & $0.0005$  \\ 
 & \multirow{4}{*}{\sparqlset} & Batch Size & $16, 32, 48$ & $48$ \\ 
 &  &  Step decay & $[0.01, 0.8]$ & $0.1$ \\ 
 &  & Step size & $1,2,5,10$ & $2$ \\  
 &  & Learning rate & $[1\times10^{-5}, 1\times10^{-2}]$ & $0.0005$  \\ 
\bottomrule

\end{tabular}
% }
\label{table:hyperparam}
\end{table*}

%% file: carmera_readay.bbl
\begin{thebibliography}{29}
\providecommand{\natexlab}[1]{#1}
\providecommand{\url}[1]{\texttt{#1}}
\expandafter\ifx\csname urlstyle\endcsname\relax
  \providecommand{\doi}[1]{doi: #1}\else
  \providecommand{\doi}{doi: \begingroup \urlstyle{rm}\Url}\fi

\bibitem[Antol et~al.(2015)Antol, Agrawal, Lu, Mitchell, Batra, Zitnick, and
  Parikh]{antol2015vqa}
Stanislaw Antol, Aishwarya Agrawal, Jiasen Lu, Margaret Mitchell, Dhruv Batra,
  C.~Lawrence Zitnick, and Devi Parikh.
\newblock Vqa: Visual question answering.
\newblock In \emph{Proc. of the IEEE international conference on computer
  vision (ICCV)}, 2015.

\bibitem[Bahdanau et~al.(2015)Bahdanau, Cho, and Bengio]{bahdanau2014neural}
Dzmitry Bahdanau, Kyunghyun Cho, and Yoshua Bengio.
\newblock Neural machine translation by jointly learning to align and
  translate.
\newblock In \emph{Proc. the International Conference on Learning
  Representations (ICLR)}, 2015.

\bibitem[Berant et~al.(2013)Berant, Chou, Frostig, and
  Liang]{berant2013semantic}
Jonathan Berant, Andrew Chou, Roy Frostig, and Percy Liang.
\newblock Semantic parsing on freebase from question-answer pairs.
\newblock In \emph{Proc. of the Conference on Empirical Methods in Natural
  Language Processing (EMNLP)}, 2013.

\bibitem[Bordes et~al.(2014)Bordes, Chopra, and Weston]{bordes2014question}
Antoine Bordes, Sumit Chopra, and Jason Weston.
\newblock Question answering with subgraph embeddings.
\newblock In \emph{Proc. of the Conference on Empirical Methods in Natural
  Language Processing (EMNLP)}, oct 2014.

\bibitem[Chakraborty et~al.(2019)Chakraborty, Lukovnikov, Maheshwari, Trivedi,
  Lehmann, and Fischer]{chakraborty2019introduction}
Nilesh Chakraborty, Denis Lukovnikov, Gaurav Maheshwari, Priyansh Trivedi, Jens
  Lehmann, and Asja Fischer.
\newblock Introduction to neural network based approaches for question
  answering over knowledge graphs.
\newblock \emph{CoRR}, 2019.

\bibitem[Cheng and Lapata(2018)]{cheng2018weakly}
Jianpeng Cheng and Mirella Lapata.
\newblock Weakly-supervised neural semantic parsing with a generative ranker.
\newblock In \emph{Proc. of the Conference on Computational Natural Language
  Learning (CoNLL)}, 2018.

\bibitem[Dong and Lapata(2016)]{li2016lang2log}
Li~Dong and Mirella Lapata.
\newblock Language to logical form with neural attention.
\newblock In \emph{Proc. the Annual Meeting of the Association for
  Computational Linguistics (ACL)}, 2016.

\bibitem[Finegan-Dollak et~al.(2018)Finegan-Dollak, Kummerfeld, Zhang,
  Ramanathan, Sadasivam, Zhang, and Radev]{catherine2018improving}
Catherine Finegan-Dollak, Jonathan~K. Kummerfeld, Li~Zhang, Karthik Ramanathan,
  Sesh Sadasivam, Rui Zhang, and Dragomir Radev.
\newblock Improving text-to-sql evaluation methodology.
\newblock In \emph{Proc. the Annual Meeting of the Association for
  Computational Linguistics (ACL)}, 2018.

\bibitem[Guo et~al.(2018)Guo, Tang, Duan, Zhou, and Yin]{NIPS2018daitoact}
Daya Guo, Duyu Tang, Nan Duan, Ming Zhou, and Jian Yin.
\newblock Dialog-to-action: Conversational question answering over a
  large-scale knowledge base.
\newblock In \emph{Proc. the Advances in Neural Information Processing Systems
  (NeurIPS)}. 2018.

\bibitem[He and Golub(2016)]{hegolub2016character}
Xiaodong He and David Golub.
\newblock Character-level question answering with attention.
\newblock In \emph{Proc. of the Conference on Empirical Methods in Natural
  Language Processing (EMNLP)}, 2016.

\bibitem[Johnson et~al.(2016)Johnson, Pollard, Shen, Li-wei, Feng, Ghassemi,
  Moody, Szolovits, Celi, and Mark]{johnson2016mimic}
Alistair~EW Johnson, Tom~J Pollard, Lu~Shen, H~Lehman Li-wei, Mengling Feng,
  Mohammad Ghassemi, Benjamin Moody, Peter Szolovits, Leo~Anthony Celi, and
  Roger~G Mark.
\newblock Mimic-iii, a freely accessible critical care database.
\newblock \emph{Scientific data}, 2016.

\bibitem[Johnson et~al.(2017)Johnson, Hariharan, van~der Maaten, Fei-Fei,
  Lawrence~Zitnick, and Girshick]{johnson2017clevr}
Justin Johnson, Bharath Hariharan, Laurens van~der Maaten, Li~Fei-Fei,
  C.~Lawrence~Zitnick, and Ross Girshick.
\newblock Clevr: A diagnostic dataset for compositional language and elementary
  visual reasoning.
\newblock In \emph{Proc. of the IEEE conference on computer vision and pattern
  recognition (CVPR)}, 2017.

\bibitem[Liang et~al.(2017)Liang, Berant, Le, Forbus, and Lao]{liang2016neural}
Chen Liang, Jonathan Berant, Quoc Le, Kenneth~D Forbus, and Ni~Lao.
\newblock Neural symbolic machines: Learning semantic parsers on freebase with
  weak supervision.
\newblock In \emph{Proc. the Annual Meeting of the Association for
  Computational Linguistics (ACL)}, 2017.

\bibitem[Luong et~al.(2015)Luong, Pham, and Manning]{luong2015effective}
Minh-Thang Luong, Hieu Pham, and Christopher~D Manning.
\newblock Effective approaches to attention-based neural machine translation.
\newblock In \emph{Proc. of the Conference on Empirical Methods in Natural
  Language Processing (EMNLP)}, 2015.

\bibitem[McCann et~al.(2018)McCann, Keskar, Xiong, and
  Socher]{bryan2018decathlon}
Bryan McCann, Nitish~Shirish Keskar, Caiming Xiong, and Richard Socher.
\newblock The natural language decathlon: Multitask learning as question
  answering.
\newblock \emph{arXiv preprint arXiv:1806.08730}, 2018.

\bibitem[Neelakantan et~al.(2015)Neelakantan, Le, and
  Sutskever]{neelakantan2015neural}
Arvind Neelakantan, Quoc~V Le, and Ilya Sutskever.
\newblock Neural programmer: Inducing latent programs with gradient descent.
\newblock In \emph{Proc. the International Conference on Learning
  Representations (ICLR)}, 2015.

\bibitem[Nguyen et~al.(2016)Nguyen, Rosenberg, Song, Gao, Tiwary, Majumder, and
  Deng]{nguyen2016ms}
Tri Nguyen, Mir Rosenberg, Xia Song, Jianfeng Gao, Saurabh Tiwary, Rangan
  Majumder, and Li~Deng.
\newblock Ms marco: a human-generated machine reading comprehension dataset.
\newblock In \emph{CoCo@ NIPS}, 2016.

\bibitem[Pampari et~al.(2018)Pampari, Raghavan, Liang, and
  Peng]{pampari2018emrqa}
Anusri Pampari, Preethi Raghavan, Jennifer Liang, and Jian Peng.
\newblock emrqa: A large corpus for question answering on electronic medical
  records.
\newblock In \emph{Proc. of the Conference on Empirical Methods in Natural
  Language Processing (EMNLP)}, 2018.

\bibitem[Rajpurkar et~al.(2016)Rajpurkar, Zhang, Lopyrev, and
  Liang]{rajpurkar2016squad}
Pranav Rajpurkar, Jian Zhang, Konstantin Lopyrev, and Percy Liang.
\newblock Squad: 100,000+ questions for machine comprehension of text.
\newblock In \emph{Proc. of the Conference on Empirical Methods in Natural
  Language Processing (EMNLP)}, 2016.

\bibitem[Saha et~al.(2019)Saha, Ansari, Laddha, Sankaranarayanan, and
  Chakrabarti]{saha2019complex}
Amrita Saha, Ghulam~Ahmed Ansari, Abhishek Laddha, Karthik Sankaranarayanan,
  and Soumen Chakrabarti.
\newblock Complex program induction for querying knowledge bases in the absence
  of gold programs.
\newblock \emph{Transactions of the Association for Computational Linguistics},
  2019.

\bibitem[See et~al.(2017)See, Liu, and Manning]{see2017get}
Abigail See, Peter~J Liu, and Christopher~D Manning.
\newblock Get to the point: Summarization with pointer-generator networks.
\newblock In \emph{Proc. the Annual Meeting of the Association for
  Computational Linguistics (ACL)}, 2017.

\bibitem[Seo et~al.(2017)Seo, Kembhavi, Farhadi, and
  Hajishirzi]{seo2016bidirectional}
Minjoon Seo, Aniruddha Kembhavi, Ali Farhadi, and Hannaneh Hajishirzi.
\newblock Bidirectional attention flow for machine comprehension.
\newblock In \emph{Proc. the International Conference on Learning
  Representations (ICLR)}, 2017.

\bibitem[Sutskever et~al.(2014)Sutskever, Vinyals, and
  Le]{sutskever2014sequence}
Ilya Sutskever, Oriol Vinyals, and Quoc~V Le.
\newblock Sequence to sequence learning with neural networks.
\newblock In \emph{Proc. the Advances in Neural Information Processing Systems
  (NeurIPS)}, 2014.

\bibitem[Wang et~al.(2020)Wang, Shi, and Reddy]{wang2020text}
Ping Wang, Tian Shi, and Chandan~K Reddy.
\newblock Text-to-sql generation for question answering on electronic medical
  records.
\newblock In \emph{Proc. the International Conference on World Wide Web (WWW)},
  2020.

\bibitem[Xu et~al.(2017)Xu, Liu, and Song]{xu2017sqlnet}
Xiaojun Xu, Chang Liu, and Dawn Song.
\newblock Sqlnet: Generating structured queries from natural language without
  reinforcement learning.
\newblock \emph{arXiv preprint arXiv:1711.04436}, 2017.

\bibitem[Yih et~al.(2015)Yih, Chang, He, and Gao]{yih2015semantic}
Scott Wen-tau Yih, Ming-Wei Chang, Xiaodong He, and Jianfeng Gao.
\newblock Semantic parsing via staged query graph generation: Question
  answering with knowledge base.
\newblock In \emph{Proc. the Annual Meeting of the Association for
  Computational Linguistics (ACL)}, 2015.

\bibitem[Yin et~al.(2018)Yin, Zhou, He, and Neubig]{yin2018structvae}
Pengcheng Yin, Chunting Zhou, Junxian He, and Graham Neubig.
\newblock Structvae: Tree-structured latent variable models for semi-supervised
  semantic parsing.
\newblock In \emph{Proc. the Annual Meeting of the Association for
  Computational Linguistics (ACL)}, 2018.

\bibitem[Yu et~al.(2018)Yu, Zhang, Yang, Yasunaga, Wang, Li, Ma, Li, Yao,
  Roman, Zhang, and Radev]{Tao2018spider}
Tao Yu, Rui Zhang, Kai Yang, Michihiro Yasunaga, Dongxu Wang, Zifan Li, James
  Ma, Irene Li, Qingning Yao, Shanelle Roman, Zilin Zhang, and Dragomir Radev.
\newblock Spider: A large-scale human-labeled dataset for complex and
  cross-domain semantic parsing and text-to-sql task.
\newblock In \emph{Proc. of the Conference on Empirical Methods in Natural
  Language Processing (EMNLP)}, 2018.

\bibitem[Zhong et~al.(2017)Zhong, Xiong, and Socher]{zhong2017seq2sql}
Victor Zhong, Caiming Xiong, and Richard Socher.
\newblock Seq2sql: Generating structured queries from natural language using
  reinforcement learning.
\newblock \emph{arXiv preprint arXiv:1709.00103}, 2017.

\end{thebibliography}
